\documentclass[manuscript]{aastex}
\slugcomment {ApJ, in press (accepted 2009 June 22)}

\shorttitle{Hydrocarbons, CO, and coronal lines in NGC~1068}
\shortauthors{}

\begin{document}

\title{The 3--5~$\mu$m Spectrum of NGC 1068 at High Angular Resolution: 
Distribution of Emission and Absorption Features across the Nuclear 
Continuum Source.}

\author{T. R. Geballe and R. E. Mason}
\affil{Gemini Observatory, 670 N. A'ohoku Place, Hilo, HI 96720}

\author{A. Rodr\'iguez-Ardila} \affil{Laborat\'orio Nacional de 
Astrof\'isica, Rua dos Estados Unidos 154, Bairro das Nac\"oes, CEP 
37500-000, Itajub\'a, MG, Brazil}

\author{D. J. Axon} \affil{Department of Physics, Rochester Institute of 
Technology, 84 Lomb Memorial Drive, Rochester, NY 14623; and School of 
Mathematical \& Physical Sciences, University of Sussex, Brighton, East 
Sussex BN1 9QH, UK}

\begin{abstract}

We report moderate resolution 3-5~$\mu$m spectroscopy of the nucleus of 
NGC~1068 obtained at 0.3\arcsec\ (20~pc) resolution with the 
spectrograph slit aligned approximately along the ionization cones of 
the AGN. The deconvolved FWHM of the nuclear continuum source in this 
direction is 0.3\arcsec. Four coronal lines of widely different 
excitations were detected; the intensity of each peaks near radio knot 
C, approximately 0.3\arcsec\ north of the infrared continuum peak, where 
the radio jet changes direction. Together with the broadened line 
profiles observed near that location, this suggests that 
shock-ionization is the dominant excitation mechanism of the coronal 
lines. The depth of the 3.4~$\mu$m hydrocarbon absorption is maximum at 
and just south of the continuum peak, similar to the 10~$\mu$m silicate 
absorption. That and the similar and rapid variations of the optical 
depths of both features across the nucleus suggest that substantial 
portions of both arise in a dusty environment just in front of the 
continuum source(s). A new and tighter limit is set on the column 
density of CO. Although clumpy models of the dust screen might explain 
the shallowness of the silicate feature, the presence of the 3.4~$\mu$m 
feature and the absence of CO are strongly reminiscent of Galactic 
diffuse cloud environments and a consistent explanation for them and the 
observed silicate feature is found if all three phenomena occur in such 
an environment, existing as close as 10~pc to the central engine.

\end{abstract}

\section{Introduction}

The unified model of active galactic nuclei (AGNs) explains the most 
basic observed properties of AGNs in terms of toroids of dust and gas, 
whose orientations to our line of sight determine the observed 
characteristics of these objects. The infrared spectral energy 
distributions (SEDs) of AGNs betray the presence of dust in their 
nuclei. Some insight into the properties and distribution of the dust 
has been gained from observations and modeling of the 9.7~$\mu$m 
silicate feature \citep{mas06,hoe07,roc06,roc07} and the 3.4~$\mu$m 
hydrocarbon absorption band \citep{dar04,mas04,mas07}. Notable findings 
include differences in the spectral profile of the 10~$\mu$m silicate 
feature between galaxies with differing amounts of obscuration 
\citep{roc07}, and evidence that the toroids are rather small, only a 
few parsecs in diameter at 10 microns 
\citep{jaf04,pac05,pon06,tri07,rab09}. At shorter wavelengths many AGNs 
are known to have nuclear point sources, whose angular extents are 
diffraction-limited (e.g., at $\sim$0.1\arcsec\ resolution as viewed in 
the $K$-band).

The dust close to an AGN is generally expected to be accompanied by 
large quantities of molecular gas \citep{kro86}, so it might be assumed 
that medium or high resolution infrared spectroscopy would result in the 
detection of molecular absorption lines there. Surprisingly, probably 
the easiest interstellar infrared molecular band to detect in 
absorption, the CO fundamental ($v$=1-0) band, whose most prominent 
lines are located at 4.5-5.0~$\mu$m, has not yet been found toward the 
nucleus of the prototypical and nearby \citep[14.4Mpc, 
1\arcsec\~=~72~pc;][]{tul88} Seyfert 2 galaxy, NGC~1068, at both 
120~km~s$^{-1}$ resolution \citep{lut04} and at 20~km~s$^{-1}$ 
resolution \citep{mas06}.  Line {\it emission} from molecular hydrogen 
in portions of the central few parsecs \citep{mul09} is the only direct 
evidence for gas phase molecules there.

A clumpy torus was first suggested by \citet{kro88} to account for dust 
surviving close to AGNs and has been invoked by \citet{nen02} to explain 
the behavior of the silicate feature with orientation and the spectral 
breadth of the far-infrared emission. The model is supported in some 
specific cases by fits to the SED of an AGN \citep[e.g.,][]{hon08}.  In 
NGC~1068 the presence of prominent dust absorption bands both to longer 
and shorter wavelengths than the CO fundamental band suggest that some 
CO absorption ought to be present. In Galactic dense cloud material, a 
silicate optical depth of 0.4, the value found by \citet{mas06} toward 
the nucleus of NGC~1068, implies an H$_2$ column density of 
0.7~$\times$~10$^{22}$~cm$^{-2}$ and a CO column density of 
1.0~$\times$~10$^{18}$~cm$^{-2}$ \citep{roc84,boh78,lee96}, which should 
be easily detectable in the fundamental band. Toward the diffuse cloud 
in front of Cyg OB2 No. 12, which has $A_{V}\sim10$ mag, the 10~$\mu$m 
silicate absorption is of comparable strength to that in NGC~1068 and 
the 3.4~$\mu$m absorption band, generally assumed to arise from 
stretching of C-H bonds in solid-state hydrocarbons \citep{san91}, is 
considerably weaker than that band in NGC~1068 
\citep[e.g.,][]{whi97,mas04}. Nevertheless several (narrow) lines 
in the CO fundamental band have optical depths of ~0.3 \citep{geb99}.

One way for CO to be present but difficult to detect is if the CO lines 
are so broad that they blend together, so that individual lines cannot 
be seen.  Using different methods \citet{geb00} and \citet{bia07} have 
each estimated the mass of the nuclear black hole in NGC~1068 to be 
$\sim$2~$\times$~10$^{7}$~M$_{\odot}$. The dimension of the outer of the 
torus-like clouds observed by \citet{jaf04} and recently re-observed by 
\citet{rab09} using mid-infrared interferometry is 2~pc. Together these 
values imply an orbital speed of 225~km~s$^{-1}$, slightly less than 
half the spacing of lines in the CO fundamental band. Thus, if absorbing 
CO is located at the outer edge of the torus, each CO transition would 
produce an absorption 450~km~s$^{-1}$ wide, which is comparable to the 
CO line spacing. In that case and assuming that the CO rotational states 
are thermalized to at least a few hundred Kelvins, the $M$ band spectrum 
would show a broad and shallow depression covering most of the band and 
would not be easily detectable in previous data sets. Any CO orbiting 
considerably farther from the nucleus, however, should be readily 
detectable in individual lines, if present with sufficient column 
density. For example, gas orbiting at a radius of 10~pc would produce 
lines with full widths at zero intensity (FWZIs) of 
$\sim$200~km~s$^{-1}$.

The high signal-to-noise ratio attainable in the $M$ band for NGC~1068, 
has allowed us to conduct a new and more sensitive search for CO 
absorption against the $M$ band continuum at the nucleus of this galaxy. 
We also have obtained a 3-4~$\mu$m spectrum of the nucleus in very good 
seeing and with high enough S/N to begin to discern the spatial 
distribution of the 3.4~$\mu$m feature within 0.4\arcsec\ (30~pc) of the 
nucleus. In the Galaxy this band, due to the C-H stretch in ethyl and 
methyl groups attached to hydrocarbons, is observed only in diffuse 
clouds \citep{pen94,chi02}. Recent observations of the spatial behavior 
of the 9.7~$\mu$m silicate feature toward AGNs have provided some 
information on the dust distribution in the central few tens of parsecs 
\citep{mas06,rhe06,roc06,roc07,you07}, in several cases revealing 
extended dusty structures perhaps associated with material fueling the 
central engine \citep{pac07}. \citet{mul09} have found stronger evidence 
for fueling of the NGC~1068 AGN, based on high angular resolution 
observations of H$_2$ line emission within the central few arc-seconds. 
More detailed observations of the 3.4~$\mu$m band and a more sensitive 
search for CO absorption promise a complementary view of the material in 
the center of NGC~1068.
 
The 3--5~$\mu$m region also includes a number of high-excitation 
forbidden emission lines. Such ``coronal" lines in AGNs have been a
subject of interest during the last few decades for several reasons. 
First, their high ionization energies ($>$100 eV) imply that they must 
be excited by shocks and/or hard photons from the AGN, with negligible 
contribution from hot stars. If the possible sources of excitation can 
be disentangled, for example through observation of the locations at 
which various lines arise, the line strengths and ratios may yield 
valuable information on the nature of the AGN continuum in otherwise 
inaccessible regions of the UV/X-ray spectrum. Second, if they are 
emitted in the inner regions of the narrow line region (NLR), or even 
associated with the putative torus of the unified model, they can be 
used indirectly to study those structures. Third, the widths, profiles 
and wavelength shifts of the lines may map extreme conditions such as 
shocks and outflows in regions close to the central engine, which may be 
spatially resolvable in nearby AGNs.

Optical spectroscopy from {\it Hubble Space Telescope} (HST) has 
provided detailed information on the morphology of the emission line 
regions and complex velocity fields of the ionized gas associated with 
radio structures close to the AGNs \citep[for NGC~1068 see][]{axo98}. 
Although efforts have been made to understand the infrared spectra of 
AGNs, because of the lack of high spatial resolution information on a 
well-defined sample, a consensus on the location and origin of the 
infrared coronal lines has not yet been reached. In this respect, 
NGC~1068 is a useful laboratory because of the prominent coronal line 
spectrum that it displays at optical, near-infrared and mid-infrared 
wavelengths. Whereas the prominent coronal lines in the optical are 
largely from Fe, the 1.0--2.5~$\mu$m region includes several other 
species, notably Si, Ca and Al, and the 3.0--5.0~$\mu$m range includes 
lines from Al, Si, Ar, and Na.

The new 3--5~$\mu$m observations and their reduction are outlined in 
\S\ref{obs}, and the strengths, distributions and profiles of the 
emission lines and molecular and solid state absorption features are 
described in \S\ref{emission} and \S\ref{absorption}, respectively. 
Comparisons with previous observations are mostly found in 
\S\ref{compare}. The implications of these findings for our 
understanding of the environments producing the coronal lines and solid 
state and possible molecular absorptions are discussed separately in 
\S\ref{discuss}.

\begin{deluxetable}{ccccc}
\tablecaption{NGC~1068 Observing Log 10 August 2006 
\label{tbl-1}}
\tablewidth{0pt}
\tablehead{
\colhead{Wavelength} & \colhead{Exp. Time} & \colhead{R} &
\colhead{NGC~1068 FWHM$^{a}$} & \colhead{HR~691 FWHM$^{a}$} \\ 
\colhead{range} & \colhead{seconds} & \colhead{$\lambda$/$\Delta$$\lambda$} 
& \colhead{arcsec} &\colhead{arcsec}
}
\startdata 
2.9--3.6~$\mu$m & 360 & 1400 & 0.41$\pm$0.02 & 0.31$\pm$0.02\\ 
3.6--4.2~$\mu$m & 360 & 2400 & 0.42$\pm$0.02 & 0.30$\pm$0.03\\ 
4.4--5.3~$\mu$m & 360 & 2000 & 0.41$\pm$0.03 & 0.29$\pm$0.03\\ 
\enddata 
\tablenotetext{a}{Measured along the slit at 3.4~$\mu$m, 3.9~$\mu$m, 
and 4.6~$\mu$m.} 
\end{deluxetable}

\begin{figure} \begin{center} 
\includegraphics[width=12cm]{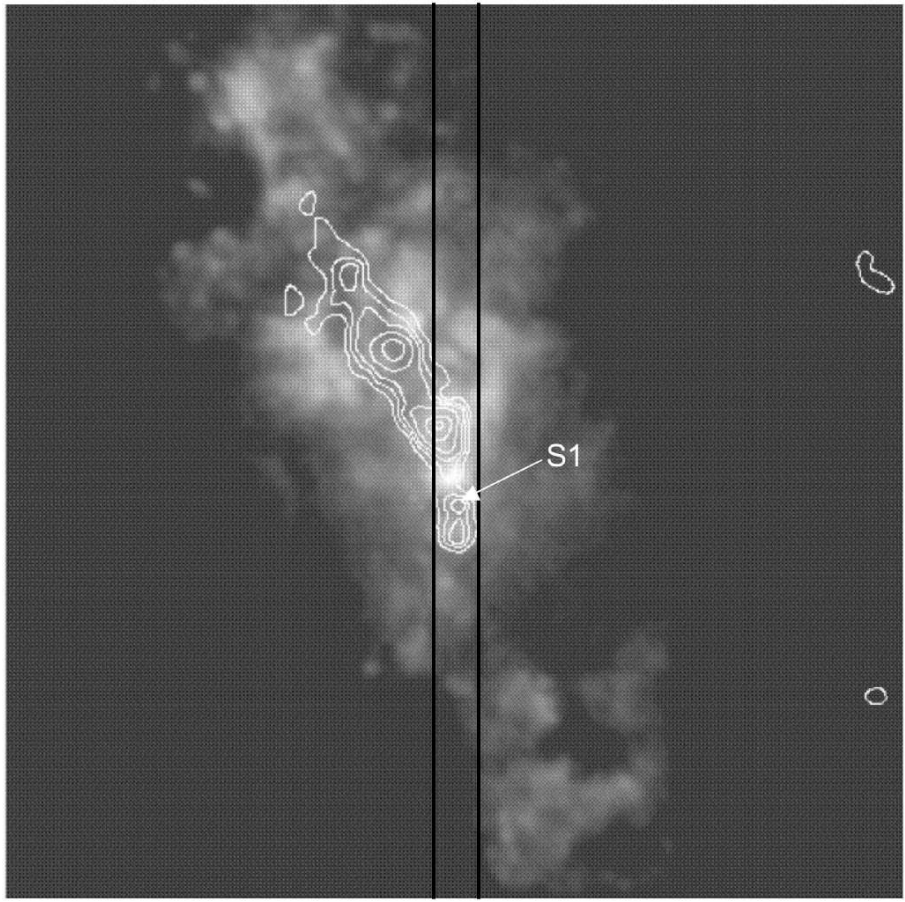}
\caption{Orientation of the 0.2\arcsec\ slit used to obtain the 
3--5~$\mu$m spectra. North is up and east is to the left. Edges of the 
slit (vertical lines) are overlayed on Fig.~1 of \citet{cap97}, a 
4\arcsec~$\times$~4\arcsec\ field containing the O~III image from 
\citet{mac94} and the 6~cm contours of \citet{mux96}. Radio source S1, 
as designated e.g. in \citet{gal04}, is coincident with the 
near-infrared continuum peak.
}
\label{fig:slit}
\end{center}
\end{figure}

\section{Observations and Data Reduction}
\label{obs}

Spectra of the nuclear region of NGC~1068 and of a nearby calibration 
star, HR 691 (F0V, L=4.7), were obtained at the United Kingdom Infrared 
Telescope (UKIRT) on Mauna Kea on 2006 August 10. The facility 
1--5~$\mu$m imager/spectrograph UIST \citep{ram04} was used, with its 
0.2\arcsec\ (2-pixel wide) slit centered on the continuum peak and 
oriented north-south, as shown in Fig.~\ref{fig:slit}.  With that slit 
width UIST delivers resolving powers, R, of 1400-2400 in various parts 
of the 3--5~$\mu$m region. The orientation roughly corresponds to the 
position angle of the ionization cones in the central 0.4\arcsec\ 
\citep{gal96}. Rows of the array are separated by 0.10\arcsec\ along the 
slit. Table~1 summarizes the observational parameters.

The observations were made in superb seeing conditions and through a 
photometric and relatively dry sky (1.5~mm of precipitable H$_2$O). 
Three spectral regions were covered, 2.9--3.6~$\mu$m, 3.6--4.2~$\mu$m, 
and 4.4--5.3~$\mu$m. Total exposure times were six minutes in each
of the above spectral intervals, and the telescope was nodded 12\arcsec\
north and back, along the slit in an ABBA pattern which was repeated
every two minutes. Long-slit spectra and imaging of H~I Br~$\gamma$ at
2.17~$\mu$m \citep{tam91, dav98} have detected no line emission at the
offset location.

HR~691 was observed in each spectral interval just prior to NGC~1068 as 
a flux calibrator and to remove telluric absorption lines from the 
spectra of NGC~1068 during data reduction.  The differences in mean 
airmass between the pairs of observations of NGC~1068 and calibration 
star were each one percent or less. The intensity profiles of the 
continuum from HR~691, measured along the slit have full widths at half 
maximum (FWHMs) of approximately three pixels 0.3\arcsec\ (equivalent to 
22~pc at the distance of NGC~1068) in all three wavelength bands 
observed. Assuming that the intensity distribution observed along the 
slit is circularly symmetric, we estimate that 55~$\pm$~10 percent of 
the signal from HR~691 fell out of the slit.

Data reduction employed Starlink Figaro routines to extract spectra in 
individual rows of the UIST detector array from coadditions of several 
subtracted pairs of images that had been flat-fielded in the UKIRT 
reduction pipeline. Following manual removal of spikes, spectra from 
pairs of adjacent rows were combined to produce spectra covering various 
0.2\arcsec~$\times$~0.2\arcsec\ regions along the slit. These were 
ratioed by the extracted spectrum of the calibration star.  Wavelength 
calibration, obtained from telluric lines in the spectrum of the 
calibration star is better than 0.0001~$\mu$m. All wavelength scales in 
this paper are {\it in vacuo} and not corrected for the systemic 
velocity of NGC~1068. The flux calibrations of the reduced 
2.9--3.6~$\mu$m and 3.5--4.1~$\mu$m spectral segments agreed to 
3.5~percent in the wavelength interval of overlap, attesting to the 
excellent weather conditions. Much of the 4.4--5.4~$\mu$m region that 
lies outside of the 4.5--5.0~$\mu$m interval is unusable or of little 
value because of low telluric transmission and is not shown here. In 
addition, we do not include the reduced spectrum in several narrow 
wavelength intervals in the 4.80--5.00~$\mu$m region, where deep 
telluric H$_2$O lines severely degrade the spectra of both the galaxy 
and the calibration star.

In the 3--5~$\mu$m interval the spectrum of the calibration star 
contains several prominent absorption lines of hydrogen in the Brackett, 
Pfund, and Humphreys series. Of these only Br~$\alpha$ falls in a 
relatively clean portion of the spectrum and can be removed by 
interpolation across the observed feature. For the other prominent H~I 
lines we employed a Kurucz R=4,000 model spectrum of Vega with line 
strengths somewhat diluted in an attempt to match the H I lines of an 
F0V star. This technique proved successful in removing the Pf~$\beta$ 
and Hu 11-6 pair at 4.65-4.67~$\mu$m but did not work well for 
simultaneous removal of the Pf~$\gamma$ and Pf~$\delta$ lines in the 
short wavelength 2.9--3.6~$\mu$m segment. Small wavelength intervals 
near those lines are omitted in the spectrum presented here.

\section{Results}

\subsection{Dimension of Nuclear Source}

Full widths at half maximum (FWHMs) along the slit were determined for 
both NGC~1068 and the calibration star from the coadded spectral images 
The results are shown in Table~1. In all three wavebands the nuclear 
continuum source in NGC~1068 had a FWHM of approximately four pixels 
along the slit (compared to three pixels for the calibration star) and 
was thus partially resolved. A simple Gaussian deconvolution, using the 
stellar FWHM as the instrumental value yields intrinsic FWHMs for 
NGC~1068 of 0.27\arcsec--0.29\arcsec\ in the NS direction (corresponding 
to a characteristic NS ``radius" for the continuum emission of 
$\sim$10~pc). This appears qualitatively consistent with the adaptive 
optics imaging near these wavelengths by \citet{mar00} as well as with 
the more recent and considerably higher resolution imaging of 
\citet{gra06}. Because the central source is partly resolved in the 
present data, it is possible to crudely investigate the spatial 
distribution of the 3.4~$\mu$m absorption feature along the slit, in 
addition to accurately determining the distributions of the emission 
lines.

\subsection{Atomic Lines}
\label{emission}

\begin{figure}
\begin{center}
\includegraphics[width=12cm]{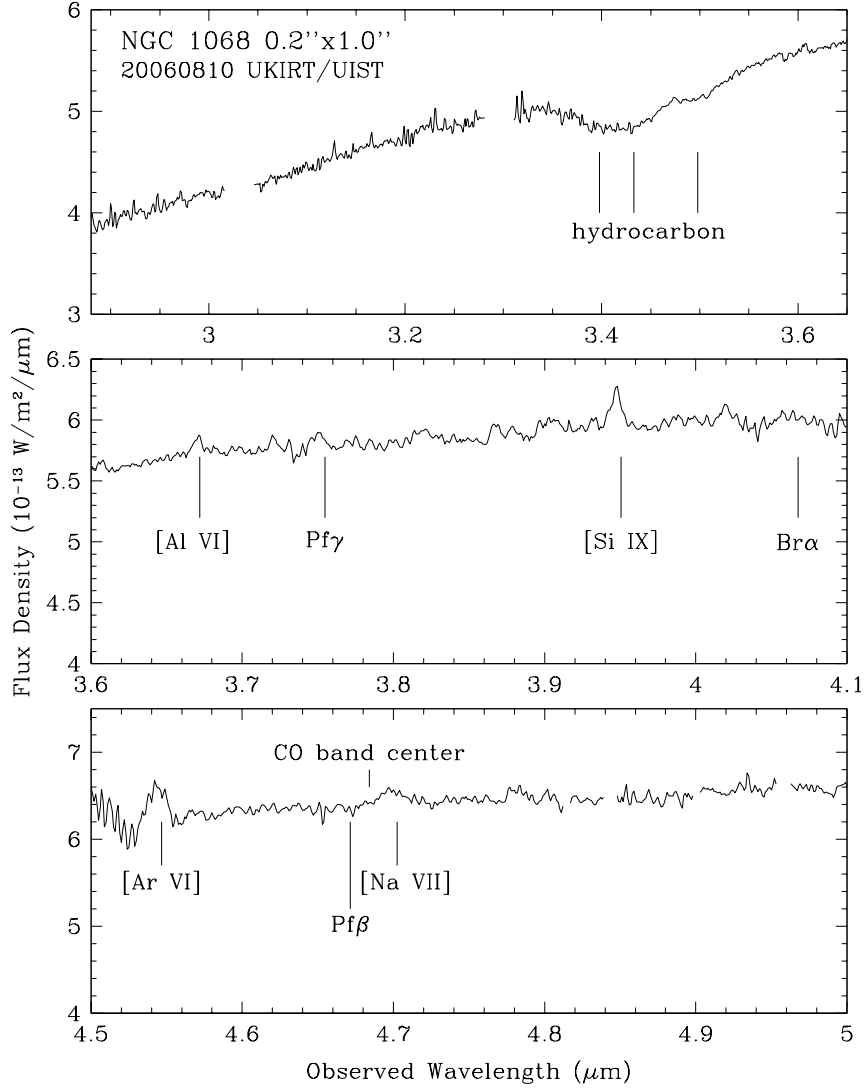}
\caption{Fig. 1: 3--5~$\mu$m spectrum of the central 0.2\arcsec\ 
$\times$ 1.0\arcsec\ (EW x NS) region of NGC~1068. Wavelengths of atomic 
lines of interest are shown, as are the the three components of the 
3.4~$\mu$m hydrocarbon feature \citep{pen94} and the band center of CO, 
all at the systemic redshift of 0.003793. The noise level varies but can 
be estimated at each wavelength from the typical point-to-point 
fluctuations of the continuum in the vicinity. For simplicity the flux 
scale assumes that both NGC~1068 and the calibration star are unresolved 
point sources and that slit losses for them are equal; this is 
approximately correct for the continuum from NGC~1068, but not for the 
much more extended line emission.}
\label{fig:spectrum}
\end{center}
\end{figure}

\begin{deluxetable}{cccc}
\tablecaption{NGC 1068 Nuclear Line Fluxes \label{tbl-2}}
\tablewidth{0pt}
\tablehead{
\colhead{Line} & \colhead{Ioniz. Energy} & \colhead{wavelength} &
\colhead{Flux in Slit$^{a}$} \\ \colhead{} & \colhead{eV} & \colhead{rest
$\mu$m} & \colhead{10$^{-16}$ W~m$^{-2}$}
}
\startdata
$[$Al\,{\sc vi}$]$ & 153.8 & 3.660 & 0.45$\pm$0.10 \\
H I 8-5 & 13.6 &  3.7405 & $<$0.3 \\
$[$Si\,{\sc ix}$]$ & 303.2 & 3.9357 & 1.5$\pm$0.15 \\
H I 5-4 & 13.6 & 4.0522 & $<$0.8 \\
$[$Ar\,{\sc vi}$]$ & 75.0 & 4.5295 & 2.5$\pm$0.7 \\
$[$Na\,{\sc vii}$]$ & 172.1 & 4.6847 & 1.1$\pm$0.3 \\
\enddata
\tablenotetext{a}{Fluxes of coronal lines are derived from 
Fig.~\ref{fig:velocity} and in the cases of [Ar\,{\sc vi}] and [Na\,{\sc 
vii}] by scaling values derived from Fig.~\ref{fig:spectrum} as 
described in the text. The upper limits to the recombination lines are 
derived from Fig.~\ref{fig:spectrum}. Upper limits and uncertainties (in 
parentheses) are 2$\sigma$; uncertainties are mainly due to uncertainty 
in the continuum level.}
\end{deluxetable}

Figure~\ref{fig:spectrum} shows the 3-5~$\mu$m spectrum extracted from 
the central ten rows (i.e., a 0.2\arcsec$\times$1.0\arcsec\ region of 
NGC~1068). The expected locations of hydrogen recombination lines and 
fine structure lines are indicated for a heliocentric systemic radial 
velocity of 1137~km~s$^{-1}$. In addition to the broad 3.4~$\mu$m 
hydrocarbon feature \citep{bri94,ima97,mas04}, four high excitation fine 
structure (coronal) lines are seen in emission: [Al\,{\sc vi}] at 
3.660~$\mu$m, [Si\,{\sc ix}] at 3.9357~$\mu$m, [Ar\,{\sc vi}] at 
4.5295~$\mu$m and [Na\,{\sc vii}] at 4.6847~$\mu$m (rest wavelengths). 
Of these the [Al\,{\sc vi}] and [Na\,{\sc vii}] lines are new 
detections. A tentative detection of the latter was reported by 
\citet{lut04}. The other lines were previously found by \citet{mar96} 
and \citet{lut00}. The coronal lines have FWZIs of over 
$\sim$1000~km~s$^{-1}$ and the peak of each line is blue-shifted 
relative to the systemic velocity of NGC~1068 by $\sim$~300~km~s$^{-1}$, 
as has been found previously for other coronal lines in NGC~1068 
\citep{mar96, lut00, ard06}. No hydrogen or helium recombination lines 
are detected in Fig.~\ref{fig:spectrum}. The marginal emission 
feature near Pf~$\gamma$ has a profile and a velocity centroid that are 
inconsistent with all of the other lines, leading us to doubt its 
reality. There is some marginally significant spectral structure in the 
vicinity of Br~$\alpha$, near 4.06~$\mu$m, but no obvious sign of line 
emission.

\begin{figure}
\begin{center}
\includegraphics[width=12cm]{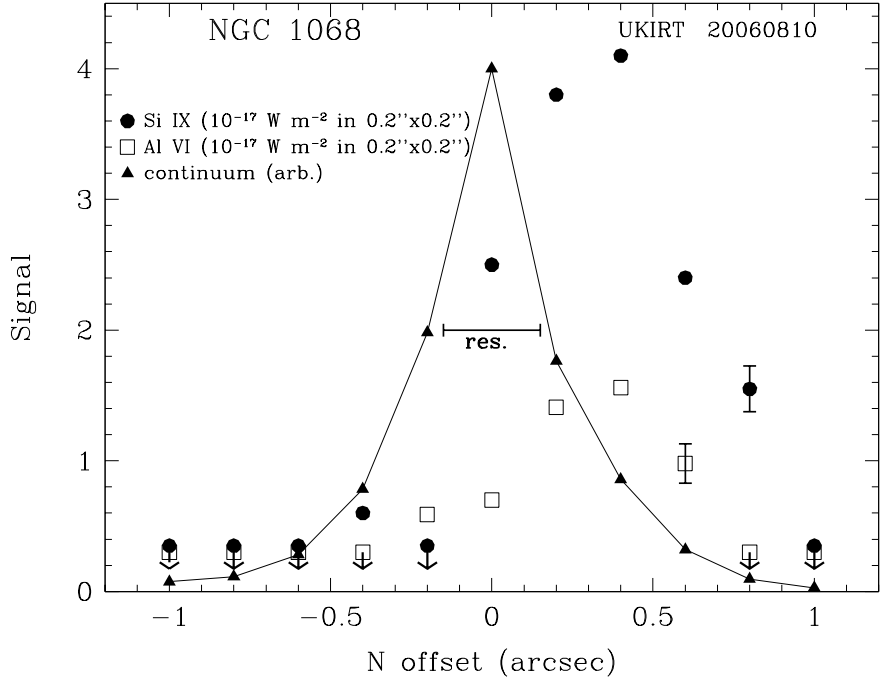}
\caption{Distribution of 3.95~$\mu$m continuum, [Si\,{\sc ix}], and 
[Al\,{\sc vi}] line flux in 0.2\arcsec\ sections and 0.2\arcsec\ 
steps along the 0.2\arcsec\ wide slit. Typical $\pm$1$\sigma$ 
uncertainties in line strengths are shown; upper limits, designated 
by downward arrows, are 2$\sigma$. The angular resolution 
(FWHM~=~0.30\arcsec), derived from the profile of the standard star, 
is indicated.
}
\label{fig:SiIX}
\end{center}
\end{figure}

The coronal line that is detected at the highest signal-to-noise ratio 
is the [Si\,{\sc ix}] line at 3.95~$\mu$m. Si$^{+8}$ has an ionization 
energy of 303~eV, making it one of the most highly excited species 
detected in NGC~1068. The next best coronal line for study in these data 
is the [Al\,{\sc vi}] line at 3.67~$\mu$m, which is more than three 
times weaker, but is situated at shorter wavelengths than other lines 
and in a region of high atmospheric transmittance so that the ambient 
background is relatively low and the noise slightly lower than at 
3.95~$\mu$m. The ionization energy of Al$^{+5}$ is 154~eV, approximately 
half that of Si$^{+8}$. Figure~\ref{fig:SiIX} shows both the continuum 
flux density and the fluxes of these two lines, in 
0.2\arcsec$\times$0.2\arcsec\ apertures along the north-south slit. The 
line fluxes peak approximately 0.3\arcsec\ north of the continuum peak 
and emission extends northward to 0.8\arcsec. Little or no line emission 
is present south of the nucleus. The peaks of the line emission from 
both [Na\,{\sc vii}] (ionization energy 172~eV) and [Ar\,{\sc vi}] 
(ionization energy 75~eV) also are offset a few tenths of an arc-second 
to the north of the continuum peak, similar to [Si\,{\sc ix}] and 
[Al\,{\sc vi}], but their detailed distributions are less certain, 
because of difficulties in accurately removing telluric absorption 
features and accurately defining the continuum level in the case of the 
[Ar\,{\sc vi}] lines, and because of the weakness of the emission in the 
case of the [Na\,{\sc vii}] line.
 
\begin{figure}
\begin{center}
\includegraphics[width=12cm]{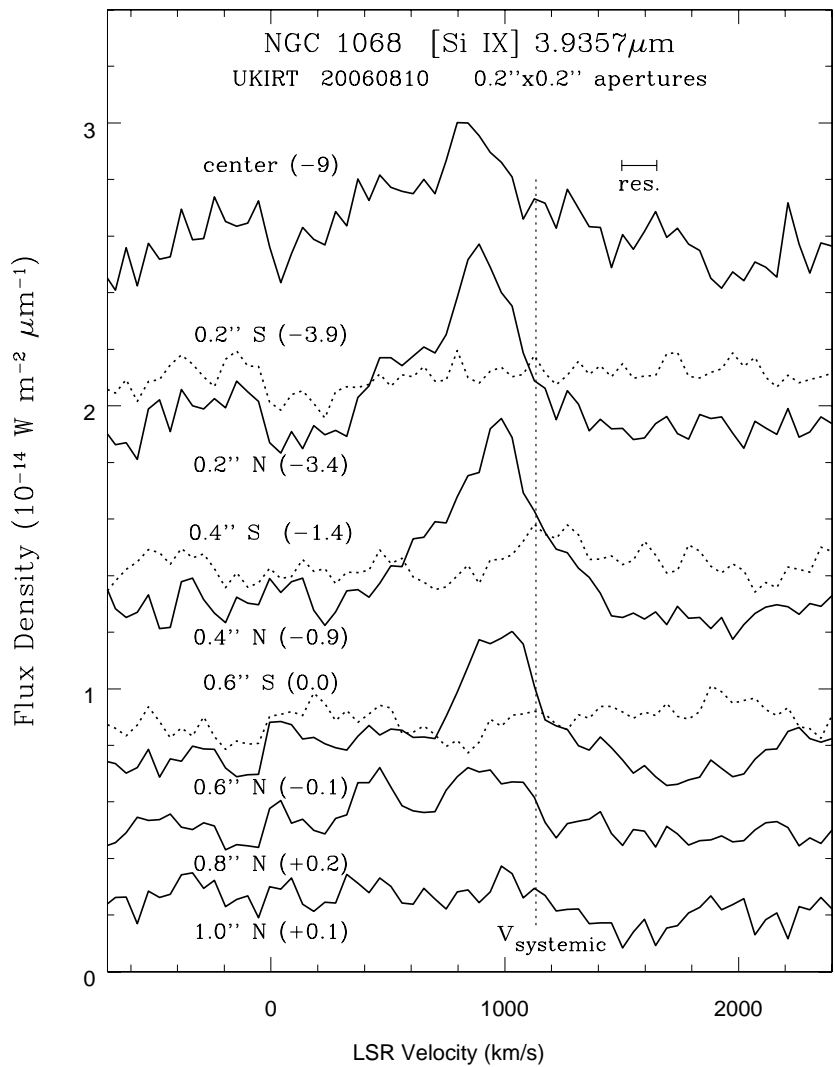}
\caption{Velocity profiles of the [Si\,{\sc ix}] line in 0.2\arcsec\ 
steps along the slit. Continuous lines denote spectra at the center and 
to the north; dashed lines indicate spectra to the south. Numbers in 
parentheses are continuum offsets. The y-axis is flux density in a 
0.2\arcsec\ $\times$~0.2\arcsec\ aperture and assumes an estimated slit 
loss factor of 2.2 for the calibration star. The systemic radial 
velocity and velocity resolution are shown.
}
\label{fig:velocity}
\end{center}
\end{figure}

Figure~\ref{fig:velocity} shows the velocity profiles of the [Si\,{\sc 
ix}] line along the north-south slit. The blue-shifted peak seen in the 
composite spectrum of Fig.~1 is present at all positions from the center 
to 0.8\arcsec\ north. At these locations the velocity of peak line 
emission is shifted between -200 and -400 km~s$^{-1}$ relative to the 
systemic velocity of NGC~1068. To the south no line emission is seen 
except perhaps 0.4\arcsec\ south where a weak line centered near the 
systemic velocity may be present. At all of the locations where the line 
is clearly detected, the velocity profile is asymmetric with an extended 
blue wing. The wing is most prominent, essentially forming a shoulder, 
in the central and 0.2~\arcsec\ north spectra. Overall the line emission 
appears to extend approximately to -1000~km~s$^{-1}$ and 
+400~km~s$^{-1}$ from the systemic velocity. This range is similar to 
that seen in other coronal lines in the 1--4~$\mu$m region, e.g., by 
\citet{lut00}.

The line fluxes within the UIST slit given in Table~2 were determined in 
various ways. Coronal line fluxes derived solely from 
Fig.~\ref{fig:spectrum} will be systematically incorrect because the 
spectrum has been flux-calibrated using a point source, but the line 
emission is more extended along the slit than is the continuum. The 
spectra in Fig.~\ref{fig:velocity} and summed fluxes for the [Si\,{\sc 
ix}] and [Al\,{\sc vi}] lines in the table have been corrected for slit 
losses, using the spectrum of the calibration star.  For [Ar\,{\sc vi}], 
and [Na\,{\sc vii}], the signal-to-noise ratios are too low to determine 
fluxes by summing their values at locations along the slit. Therefore we 
have assumed similar distributions for these lines as [Si\,{\sc ix}] and 
[Al\,{\sc vi}]. Then their fluxes in the slit can be estimated 
by scaling the fluxes derived from Fig.~\ref{fig:spectrum} by 0.60, the 
ratio of the fluxes of the [Si\,{\sc ix}] and [Al\,{\sc vi}] lines in 
Table~2 to the values for them derived from Fig.~\ref{fig:spectrum}. 
For Br~$\alpha$ and Pf~$\gamma$ upper limits were derived solely from 
the spectrum in Fig.~\ref{fig:spectrum}.

\begin{figure}
\begin{center}
\includegraphics[width=12cm]{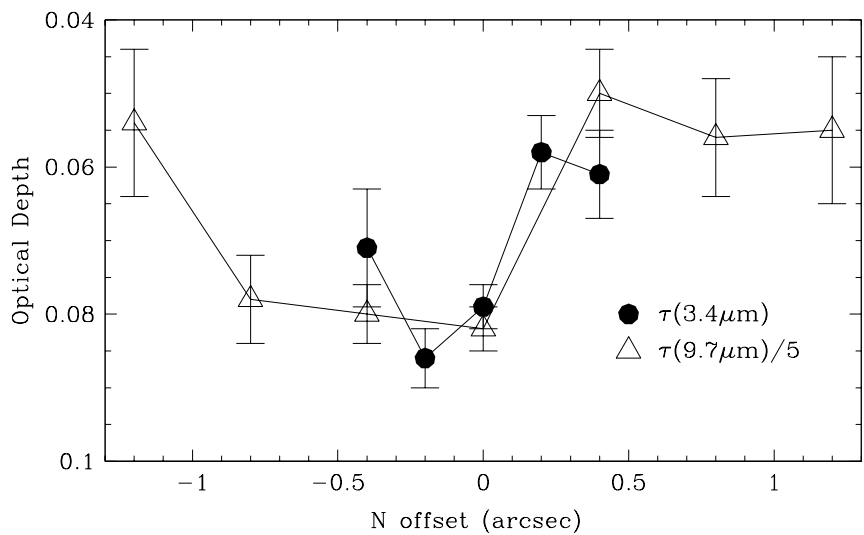}
\caption{Optical depth of the 3.4~$\mu$m hydrocarbon absorption feature
(this paper) and the 9.7~$\mu$m silicate absorption feature \citep{mas06}
across the nucleus of NGC~1068. Error bars are $\pm$1$\sigma$.
}
\label{fig:chsil}
\end{center}
\end{figure}

\subsection{Solid-state and molecular absorptions} 
\label{absorption}

The 3.4~$\mu$m hydrocarbon absorption feature is the only solid-state 
feature present in the 3--5~$\mu$m spectrum. It extends from 3.35~$\mu$m 
to perhaps 3.57~$\mu$m with a broad absorption maximum near 3.41~$\mu$m 
and a shoulder that begins at 3.47~$\mu$m. Assuming that the 
contribution of the 3.43~$\mu$m component drops to zero at 3.50~$\mu$m, 
the absorption maximum of the shoulder feature occurs close to 
3.50~$\mu$m. After accounting for the redshift of NGC~1068, the above 
wavelengths and the overall profile are in good agreement with those 
observed for the 3.4~$\mu$m feature in Galactic sources \citep{pen94}.

The optical depth of the 3.4~$\mu$m feature has been determined by 
fitting a linear continuum across the 3.10--3.35~$\mu$m interval 
(excluding the 3.29--3.33~$\mu$m region) and the 3.57--3.60~$\mu$m 
interval.  The maximum optical depth of the 3.4~$\mu$m feature shown in 
Fig.~\ref{fig:spectrum} is 0.072~$\pm$~0.004.

Using the same technique we also have measured the optical depth of the 
3.4~$\mu$m feature in 0.2\arcsec\ steps within $\pm$0.4\arcsec\ of the 
infrared continuum peak. The results are shown in Fig.~\ref{fig:chsil}, 
together with the measurements of the optical depth of the 9.7~$\mu$m 
silicate absorption from \citet{mas06} measured in 0.4\arcsec\ steps 
steps over a wider region.  At and south of the continuum peak the 
optical depth of the 3.4~$\mu$m feature is approximately 0.08, while at 
the northern locations the depth is noticeably less, approximately 0.06. 
This pattern is similar to that seen in the silicate feature. The 
implications of this are discussed in \S\ref{discuss}.

The 4.5--5.0~$\mu$m portion of the spectrum allows a new search for the 
presence of the fundamental band of carbon monoxide (CO), which should 
manifest itself as individual absorption lines if the CO line width is 
less than the line spacing ($\sim$500~km~s$^{-1}$ or as broad 
depressions on either side of the band center if the CO line profiles 
are comparable to or broader than the spacing.  \citet{mas06}, observing 
at a resolution of 20~km~s$^{-1}$, did not detect narrow CO lines. As 
discussed earlier CO absorption lines arising in gas orbiting the AGN 
outside the characteristic radius of the continuum emitting region 
should be no wider than 200~km~s$^{-1}$ (FWZI).

No signs of absorption by individual lines can be seen in the current 
spectrum, and any broad depression between the band center and the long 
wavelength cannot be deeper than about 1\%\ of the continuum. Close 
examination of the spectrum shows that the continuum in the 
4.8--4.9~$\mu$m portion of the spectrum, corresponding to P branch 
transitions with 10$<$$J$$<20$, may be depressed by roughly that amount 
relative to adjacent portions of the spectrum. This could be interpreted 
as evidence for absorbing CO in dense gas at temperatures of several 
hundred degrees and with very high velocity dispersion. For an 
isothermal absorbing slab, lines of CO at this temperature with FWHM's 
of 500~km~s$^{-1}$ and central depths of 1\%\ would produce such a 
depression and would imply a CO column density of 
$\sim$~1~$\times$10$^{17}$~cm$^{-2}$ and a gas particle density of order 
10$^{7}$~cm$^{-3}$. The corresponding portion of the R branch of CO is 
not available for comparison due to telluric absorption and 
contamination from the [Ar\,{\sc vi}] line. Because the evidence for 
this broad CO absorption is marginal we regard the above column density 
as an upper limit.

\section{Comparisons and Constraints}
\label{compare}

\subsection{Atomic lines}

Both the offsets of the peaks of the [Si\,{\sc ix}] and [Al\,{\sc vi}] 
lines from the peak of the continuum and the extents of the line 
emission are somewhat less than those reported by \citet{mar96} for the 
coronal lines that they measured along the ionization cone at position 
angle 31 degrees. However, their measurements were made at roughly five 
times lower angular resolution than those here. Thus the present 
determination of the offset should be more accurate. The distributions 
of the other two coronal lines along the NS slit are not as well known 
as those of [Si\,{\sc ix}] and [Al\,{\sc vi}], but they appear to be 
similar.

\citet{tho01} imaged the 1.96~$\mu$m coronal [Si\,{\sc vi}] line at 
0.2\arcsec\ resolution, finding it to originate mainly to the 
north-northeast of the the continuum source. The current measurements of 
the [Si\,{\sc ix}] line are largely consistent with \citet{tho01}, but 
failed to detect the fainter line emission that their data show to the 
south.  \citet{pri05} imaged the [Si\,{\sc vii}] 2.48~$\mu$m line 
emission at 0.10\arcsec\ resolution. Their figures show it to be mostly 
in a clumpy structure centered on the continuum peak. This appears to 
differ markedly from the distributions of both the [Si\,{\sc vi}] line 
imaged by \citet{tho01} and the [Si\,{\sc ix}] line observed 
spectroscopically by \citet{mar96} and by us.  The accuracy of the 
co-registration of their images is not available.  In the present data 
the lines and the continuum were measured simultaneously and the result 
is clear. The ionization energy of [Si\,{\sc vii}] lies between those of 
[Al\,{\sc vi}] and [Si\,{\sc ix}]; thus it would be surprising if the 
intensity distribution of the [Si\,{\sc vii}] line relative to that of 
the adjacent continuum were not similar to the lines observed here.

The new measurements of these lines put significant constraints on the 
size of the high-ionization line emitting region. The detected [Si\,{\sc 
ix}] flux in the UIST slit is about 3.5 times smaller than that reported 
by \citet{lut00} using the Infrared Space Observatpry (ISO) with a 
14\arcsec\ $\times$ 20\arcsec\ aperture, and three times smaller than 
that reported by \citet{mar96}, who used a 4.4\arcsec\ slit width.  
However, assuming that the [Si\,{\sc ix}] line-emitting region through 
which our observations have sliced is roughly circularly symmetric, the 
total flux in the [Si\,{\sc ix}] region is 
$\sim$5.3~$\times~10^{-16}$~W~m$^{-2}$, in agreement with the flux 
measured by ISO and, within the uncertainties, with that by Marconi et 
al. Using adaptive optics imaging \citet{pri05} fund a rather clumpy 
distribution of the [Si\,{\sc vii}] 2.48~$\mu$m line emission extended 
over $\sim$1\arcsec\, but one that does not prefer any particular 
position angle, so the assumption of circular symmetry, while likely 
incorrect in detail, should allow a reasonable estimate of total line 
flux.

Following the same line of reasoning, the [Al\,{\sc vi}] line, which has
a very similar flux distribution along the UIST slit as [Si\,{\sc ix}]
(see Figure~\ref{fig:SiIX}), would have a total flux of
1.6$\times10^{-16}$~W~m$^{-2}$. The total fluxes in the [Ar\,{\sc vi}],
and [Na\,{\sc vii}] lines would be 9~$\times$~10$^{-16}$~W~m$^{-2}$ and
4~$\times$~10$^{-16}$~W~m$^{-2}$, respectively. \citet{lut00} found a
1.7 times higher flux for the [Ar\,{\sc vi}] line, suggesting that the
distribution of this ion, which possesses the lowest ionization energy
of the four detected, is somewhat more extended than [Si\,{\sc ix}] or
possesses a large and very faint halo. \citet{pri05} and \citet{ard06}
found that the [Si\,{\sc vi}] and [Si\,{\sc vii}] lines (with ionization
energies of 167~eV and 205~eV, respectively) each have halos extending
to a radius of 70~pc (1.0\arcsec), but they do not provide the relative
contributions of the core and halo.

The absence of hydrogen 5$-$4 recombination line emission is another 
noteworthy aspect of the 3-5~$\mu$m spectrum in 
Figure~\ref{fig:spectrum}. The Br~$\gamma$ line emission detected by 
\citet{dav98} from the circum-nuclear ring of H~II regions at a radius 
of 15-16\arcsec\ has 1--2 orders of magnitude lower surface brightness 
than that seen at the nucleus by \citet{tam91}. Thus the weakness of the 
recombination line emission in Figure~\ref{fig:spectrum} is not likely 
to be due to subtraction of bright line emission in the offset beam, 
located 12\arcsec\ north of the nucleus, just inside the ring. 
\citet{dav07} have mapped the Br~$\gamma$ line emission in the vicinity 
of the nucleus at high angular resolution and have kindly made their 
data available to us. Our limit for the Br~$\alpha$ flux in the central 
0.2\arcsec~$\times$1.0\arcsec of $<$8~$\times$10$^{-17}$~W~m$^{-2}$ is 
75 times the Br~$\gamma$ line flux in the same aperture. Assuming Case B 
conditions this limit corresponds to an upper limit on the reddening of 
3.5 mag between the wavelengths of the two lines, and, for a 
$\lambda$$^{-1.8}$ dependence for the extinction, a visual extinction of 
no more than $\sim$60 mag to the Br~$\gamma$ line-emitting region.

\subsection{Solid state and molecular absorptions}

The optical depth of the 3.4~$\mu$m feature reported here is in good 
agreement with the values measured by \citet{bri94} and \citet{mas04} in 
larger apertures, but disagrees with values about twice as high reported 
by \citet{ima97} and \citet{mar03}. As the bulk of the continuum 
emission comes from the central point source, the differences with 
\citet{ima97} and \citet{mar03} are difficult to account for in terms of 
positional variations.  In the case of the former paper, incorrect 
location of the short wavelength continuum may be responsible for the 
discrepancy, but this does not seem to be the explanation for the 
\citet{mar03} result. Neither we, nor any other of the above observers 
have detected the nuclear 3.3~$\mu$m (PAH) emission feature reported by 
\citet{mar03}. It is possible that some of the features that they see 
are spurious, perhaps resulting from incorrect cancellation of strong 
telluric lines present in the 3.25--3.45~$\mu$m interval.

\citet{aya92} and \citet{lut04} searched for lines of the CO fundamental 
at resolutions of 250~km~s$^{-1}$ and 120~km~s$^{-1}$, respectively. The 
latter authors set a 3$\sigma$ upper limit of 7\% of the continuum for 
the depths of individual CO lines. In the 4.57-4.77~$\mu$m portion of 
the current spectrum, obtained at 150~km~s$^{-1}$ resolution and also 
devoid of CO lines, the signal-to-noise ratio on the continuum is about 
three times higher. Thus, the limits both on individual lines separated 
from one another (i.e., with widths less than $\sim$500~km~s$^{-1}$) and 
on a broad depression due to blended CO lines are significantly more 
stringent than those of \citet{lut04}.

Finally, \citet{aya92} reported the detection toward the nucleus of
NGC~1068 of a narrow absorption feature at 4.69~$\mu$m, which they
attributed to solid CO. Our spectrum shows no such feature, to a limiting
absorption depth of a few percent of the continuum.

\section{Discussion}
\label{discuss}

\subsection{Excitation mechanisms for coronal line emission}

One of the most important questions concerning coronal lines in AGN is 
the excitation mechanism that drives their emission. Here we discuss 
whether photoionization or shock ionization can account for the 
locations and strengths of the coronal line emission in NGC~1068. A 
summary is provided at the end of this section.

\subsubsection{Photoionization}

\citet{fer97} carried out a number of illustrative photoionization 
simulations in an effort to identify the optimal conditions and 
locations in which the coronal lines form around AGN. Assuming 
ultraviolet radiation from the central engine as the only excitation 
mechanism and plane-parallel, constant density slabs of gas, they 
determined the distances from the ionizing source in which the lines are 
emitted as a function of density. They provide line equivalent widths in 
the density-distance plane, indicating where the bulk of the emission 
occurs for each line, and allowing rough comparisons between observed 
and predicted emission line flux ratios.  The [Al\,{\sc vi}], [Ar\,{\sc 
vi}] and [Si\,{\sc ix}] lines observed here are included in their 
calculations. Their figures assume an ionizing ultraviolet continuum 
similar to that of a typical Seyfert galaxy with $L_{\rm ion} = 
10^{43.5}$ erg\,s$^{-1}$, but distances scale as $L_{\rm ion}^{1/2}$. 
\citet{pie94} and \citet{bla97} estimate ionizing luminosities 1--2 
orders of magnitude higher in NGC~1068.

In the discussion below we assume an ionizing luminosity of 
10$^{45}$~erg\,s$^{-1}$, which implies locations and dimensions of the 
emission regions are five times larger than those shown by \citet{fer97}, if 
all other parameters remain constant. The model then predicts that the 
distance and density for peak [Si\,{\sc ix}] line emission cover 
1.5--100~pc and 10$^{3}$--10$^{6.25}$~cm$^{-3}$, respectively (with the 
largest distances corresponding to the lowest densities). The observed 
distance of 20 pc (0.3\arcsec) from the nucleus to the peak line 
emission corresponds to $n$$\sim$10$^{4}$~cm$^{-3}$). If the actual peak 
is closer to the AGN than observed but obscured by dust, then higher 
density gas would be required.

For the [Al\,{\sc vi}] and [Ar\,{\sc vi}] lines the adjusted 
\citet{fer97} models predict distances to the peak of roughly 15--400~pc 
for the same density range. Although the spatial distribution of the 
[Ar\,{\sc vi}] line flux is less precise than the other lines, it 
appears to be similar to that of the other two lines, with a peak at the 
same location. Clearly the observation that the offset from the AGN to 
the peaks of these lines of with widely varying excitations are the same 
is incompatible with an ionized medium of uniform density, unless the 
extinction is highly non-uniform across the nucleus. The current 
observations can be roughly reconciled with a slab of density several 
times 10$^{4}$~cm$^{-3}$ if there is increasing extinction of the 
coronal line emission closer to AGN, of sufficient strength to mask the 
true peak of the [Si\,{\sc ix}] line emission. If the coronal line 
emission region is symmetrically distributed to the north and south of 
the AGN, the extinction would need increase further to the south of the 
AGN in order to block the line emission there from view. Evidence for 
higher extinction to the south is in fact widespread \citep[e.g.,][and 
references therein]{mas06,das06}. A rough lower limit to the additional 
extinction required to the south based on the assumption of a symmetric 
distribution of coronal line emission is two magnitudes at 3.93~$\mu$m, 
which corresponds to approximately 70 visual magnitudes. If the highly 
non-uniform extinction postulated above is not present, only a density 
distribution of ionized gas strongly peaked 20~pc north of the AGN could 
result in photoionization producing the observed maximum line emission 
from all of these ions at the same location.

The calculations of \citet{fer97} also indicate that for these ions the 
regions of significant coronal line emission can extend to much larger 
distances ($\sim$100~pc) if gas densities decrease gradually with 
distance from the AGN. The near agreement of our estimates of the 
total flux from the [Si\,{\sc ix}] and [Ar\,{\sc vi}] lines with 
measurements made by others in much larger apertures, imply that the 
ionized gas density must drop steeply at distance greater than 
$\sim$50~pc from the AGN.

\subsubsection{Shock ionization}

Evidence supporting shock ionization in the narrow line regions of AGNs 
abounds \citep{mar96,con02,pri05,ard06}. In the case of NGC~1068, 
\citet{axo98} found that 1.5\arcsec\ north of the nucleus [Fe\,{\sc 
vii}] is enhanced at the location of the radio jet. They interpreted 
this as evidence of interaction between the jet and the surrounding 
medium. The radio jet proceeds outward from the nucleus for an angular 
distance of several arc-seconds, first nearly due north and then at 
position angle 35~degrees \citep{gal96,gal04}. More recently, 
\citet{das06} and \citet{das07} have studied the influence of the radio 
jet on the kinematics of [O\,{\sc~iii}]~$\lambda\lambda$4959,5007. They 
observed an increase in the radial velocity roughly proportional to 
distance from the nucleus followed by a linear decrease beyond 
$\sim$100~pc (1.4\arcsec) and concluded that the magnitudes of these 
changes were so large that the neither can the jet be the sole driving 
force nor can gravity be the sole decelerating mechanism (frictional 
forces must be significant). However, in the case of that line the 
deceleration occurs well outside of the region where we observe the 
higher excitation 3--5~$\mu$m coronal lines.

The line emission reported here originates in ionized regions much 
closer to the nucleus than those observed by \citet{axo98} and 
\citet{das07}. Nevertheless, evidence for shock excitation of the 
coronal lines is also readily apparent when one compares the 5~GHz image 
in Fig.~1 of \citet{gal04} with our data. Assuming, as they and 
\citet{roy98} have inferred, that source S1 is coincident with the AGN, 
and hence with the infrared continuum peak, the peaks of [Si\,{\sc ix}] 
and [Al\,{\sc vi}] line emission at 0.3\arcsec\ north are coincident 
with the bright radio knot C, where the jet bends to the northeast. This 
is highly suggestive of an interaction between the jet and gas in the 
NLR. Also suggestive of an interaction at this location is a bright knot 
of H$_{2}$ 2.12~$\mu$m line emission found by \citet{mul09} connecting 
two streamers of line emission, one of which they interpret as feeding 
the innermost few parsecs.

The velocity profiles of the of [Si\,{\sc ix}] line along the spatial 
direction, shown in Figure~\ref{fig:velocity}, also appear to support a 
jet-NLR interaction. At 0.2$\arcsec$ north, a prominent blue shoulder is 
evident in the line profile. That structure also appears to be present 
0.4\arcsec\ north, but it weakens significantly at 0.6\arcsec. The 
shoulder at 0.2--0.4\arcsec\ could be associated with ionized gas 
accelerated toward us by the interaction of the jet with the ambient gas 
near knot C. The line profile variations are somewhat reminiscent of the 
behavior of the [Fe\,{\sc vii}]~$\lambda$6087 line reported by 
\citet{ard06}, but the latter is at a much larger distance north of the 
AGN and, also unlike the infrared lines, far from where that line has 
its peak intensity.

If shocks are the dominant excitation mechanism for these coronal lines 
at radio knot C, then it seems likely that some of the remaining 
[Si\,{\sc ix}] and [Al\,{\sc vi}] line flux outside of that knot 
originates in other localized regions and is associated with other radio 
knots. The knot labeled ``NE'' in Fig.~1 of \citet{gal04}, located 
0.4$\arcsec$ to the NE of knot ``C'' and not included in our slit, is a 
possible candidate for some of this missing line flux. However, judging 
from the large extent over which [Si\,{\sc vii}] was observed by 
\citet{pri05}, we do not think that shocks totally dominate over 
photoionization, and expect that some contribution of the latter is 
present, perhaps even at radio knot C.

\subsubsection{Summary}

The infrared coronal line emission peaks 20~pc to the north of the AGN 
for ions of a wide range of excitation, and extends to 50~pc in the 
highly excited [Si\,{\sc ix}] line and presumably at least as far in the 
other lines. To the south there is no clear detection of any of the 
3--5~$\mu$m lines. Increased extinction to the south can account for the 
north-south asymmetry, regardless of the line excitation mechanism. 
However, photoionization by the AGN producing peaks at equal northern 
distances from the AGN for such a large range of excitations requires a 
highly asymmetric gas distribution and/or a highly asymmetric 
distribution of obscuring dust to the north. This explanation appears 
contrived. Evidence for collisional excitation of these lines is more 
compelling; it consists of the spatial coincidence between radio knot C 
(where the jet changes direction), a prominent knot of H$_{2}$ line 
emission, and the emission peaks of all of the 3--5~$\mu$m coronal 
lines. The presence at the same location of a localized prominent high 
velocity blue-shifted shoulder on the [Si\,{\sc ix}] line provides 
further support. Higher angular resolution studies of the coronal line 
emission would allow a more detailed comparison with the morphology of 
the radio emission.

\subsection{The 3.4~$\mu$m absorption feature and the non-detection of CO}

In his subsection we discuss that portion of the interstellar medium 
producing the hydrocarbon and silicate absorption features at 3.4~$\mu$m
 and 9.7~$\mu$m, but little or no absorption by gaseous CO at 
4.5--5.0~$\mu$m. Our discussion makes use of the silicate absorption 
feature as reported by \citet{mas06}; its similar behavior across the 
nucleus as the 3.4~$\mu$m feature provides a strong clue as to the 
nature of the interstellar medium. The subsection on CO examines a 
number of possible explanations for its absence, and results in a 
similar conclusion regarding the ISM. A summary is provided at the end 
of this section.

It is important to note that in the direction along the (NS) slit the 
continuum source was partially resolved at UKIRT and therefore we are 
assured that it has a characteristic dimension of a few tens of parsecs. 
Thus the bulk of the material producing the 3.4~$\mu$m absorption 
feature and not producing CO absorption lies 10~pc or more from the 
central engine. Dust and gas in the central few parsecs can only be a 
minor contributor to the present spectrum and our arguments as to the 
nature of the interstellar medium therefore apply to the more extended 
region and not to the ``core" or the innermost region of the torus 
observed via interferometry.

\subsubsection{The 3.4~$\mu$m band}
\label{hydroc}

The spectral profile of the 3.4~$\mu$m feature is well characterized in 
the current data. Overall the profile is smooth, showing only the two 
principal sub-features seen in Galactic sources, those at 3.420~$\mu$m 
and 3.485~$\mu$m. The profile in NGC~1068 contains no evidence of a 
distinct 3.385~$\mu$m sub-feature, which is sometimes present in 
Galactic sources and sometimes blended with the 3.420~$\mu$m feature 
\citep{pen94,chi02}.

The data also reveal that the depth of the 3.4~$\mu$m feature has a 
similar north-south variation across the nucleus as the 9.7~$\mu$m 
silicate absorption (Fig.~\ref{fig:chsil}). Both features are at maximum 
depth on and just to the south of the nucleus. The similarity is also 
quantitative; the fractional increase in the depth of each feature at 
its maximum compared to the depth at adjacent locations is $\sim$40 
percent. {\it This indicates that the carriers of the features are 
significantly mixed.} The compactness and nuclear location of the region 
of enhanced 3.4~$\mu$m and silicate extinction suggests that the extra 
absorption to the south occurs quite close to the nucleus. In the 
silicate feature the weaker off-nuclear absorption extends over 
2\arcsec\ to the north and south. Because the 3~$\mu$m continuum source 
is much more compact, the depth of the 3.4~$\mu$m feature could not be 
measured more than 0.4\arcsec\ distant from the nucleus. The lower level 
and more extended silicate feature largely arises in a circumnuclear 
region of much larger dimensions than the torus-like structures seen 
with mid-IR interferometry. It may be associated with material in the 
disk of the host galaxy, which crosses the line of sight south of the 
nucleus. The distinction between these two regions is not necessarily 
sharp \citep{pac07}.

The ratio of the optical depths of the 3.4~$\mu$m and silicate features 
at the infrared continuum peak of NGC~1068 is 0.19~$\pm$~0.02. In the 
Galaxy the silicate feature is found in both diffuse and dense 
interstellar clouds, whereas the hydrocarbon feature is found only in 
diffuse clouds. In diffuse clouds far from the Galactic center the ratio 
of optical depths of the two features can be derived from their optical 
depths relative to $A_V$, $A_V$/$\tau_{9.7}$~$\approx$~18.5 
\citep{roc84} and $A_V$/$\tau_{3.42}$~$\approx$~250 \citep{pen94}, 
yielding $\tau_{3.4}$$/$$\tau_{9.7}$~=~0.07~$\pm$~0.01, a value roughly 
forty percent of that measured in NGC~1068. Toward the Galactic center, 
where $A_V$~$\approx$~30~mag), the value of $\tau_{3.4}$/$\tau_{9.7}$ 
averaged over sources observed by both \citet{chi02} and \citet{roc85} 
is 0.064~$\pm$~0.017, the same as the above value to within the 
uncertainties. The ratio {\it in just the diffuse gas on the Galactic 
center sightline} may be estimated as follows. \citet{pen94} find 
$A_V$/$\tau_{3.4}$~=~156~$\pm$16 and \citet{roc85} find 
$\tau_{9.7}$~=3.6~$\pm$~0.3. However, one-third of the visual extinction 
to the Galactic center is believed to occur in dense clouds, located 
mostly in intervening spiral arms \citep{whi97}. Following 
\citet{chi07}, for values of $A_V$$<$12~mag the ratio of visual 
extinction to silicate optical depth is the same in dense clouds as in 
diffuse clouds, and thus we associate those 10~mag of dense cloud visual 
extinction with a silicate optical depth of 0.54. This gives 
$\tau_{3.4}$/$\tau_{9.7}$~$\approx$~0.23/3.06~=~0.075 for Galactic 
center diffuse clouds, again far less than the ratio toward the nucleus 
of NGC~1068. Even if half of the silicate optical depth toward the 
Galactic center arises in the dense clouds in the Galactic disk, the 
above ratio would only be 0.15, less than the value in NGC~1068.

Clearly then, the 3.4~$\mu$m absorption in NGC~1068 is unusually strong 
relative to the silicate absorption. This cannot be explained away by 
placing the 3.4~$\mu$m absorber in a more remote location than the 
silicate feature, because even in Galactic diffuse clouds the optical 
depth of 0.08 observed in NGC~1068, corresponds to a silicate optical 
depth $\gtrapprox$1, significantly larger than is observed. The high 
ratio could be the result of the silicate absorption being significantly 
filled in, as might arise if thermal emission close to the AGN were 
absorbed in the cooler outer portions of the source. Because such 
filling in would be much less for the 3.4~$\mu$m feature, the similar 
variation in optical depths of the features across the nucleus tends to 
argue that this is not the entire explanation, however. As suggested by 
\citet{ima00}, the high ratio could be explained, at least in part, 
by the typical continuum surface at 3.4~$\mu$m lying interior to the 
typical surface at 9.7~$\mu$m such that the added pathlength, 
unobservable at 9.7~$\mu$m, contains a significant fraction of the 
observed column density of the 3.4~$\mu$m absorber. The result also 
could be a consequence of either unusual elemental abundances or the 
chemistry of the dust in the vicinity of the AGN.

\subsubsection{The non-detection of CO}

\label{co}

The CO fundamental band has been searched for previously without success 
in NGC~1068 at spectral resolutions ranging from 20~km~s$^{-1}$ to 
250~km~s$^{-1}$ \citep{aya92,lut04,mas06}. The current data, obtained at 
150~km~s$^{-1}$ resolution, significantly tightens the limit on 
absorption by CO in the warm and presumably rapidly moving gas just 
outside of the 3--5~$\mu$m continuum-emitting dust, whose FWHM along the 
ionization cones is $\sim$20~pc. In its lack of detectable CO in this 
band NGC~1068 is similar to the five other Type 2 Seyferts studied by 
\citet{lut04}, although the limits on CO absorption in those galaxies 
are considerably less strict.

We consider three possible explanations for the lack of strong CO 
absorption lines toward the AGN in NGC~1068: (1) the molecules are 
distributed among many rotational states; (2) the molecules reside in a 
clumpy obscuring medium; and (3) there is a paucity of ``dense cloud" 
molecular gas in front of the AGN. To explain the non-detection of 
transitions of CO, OH and H$_{2}$CO lines from low rotational levels 
towards the highly obscured nucleus of Cygnus~A at radio wavelengths, it 
has been pointed out that at the temperatures of the nuclear clouds the 
lowest rotational levels may contain only a small fraction of the 
molecules \citep{bar94,mal94,con95}. For example, in LTE at 300~K 
roughly twenty rotational levels of CO have populations within a factor 
of ten of that of the most populated level, whereas at 25~K only the six 
lowest levels are so highly populated. It also has been suggested by 
these authors that under some circumstances a strong and compact nuclear 
radio source could redistribute the rotational level populations in the 
surrounding gas, further weakening the strengths of lines from the low 
rotational levels. \citet{imp06} find this explanation unlikely, 
however, because the 19\% of their sources in which high-excitation OH 
lines are detected also show absorption from the ground state 
transition.

In the case of NGC~1068, whose 3--5~$\mu$m continuum-emitting dust, and 
presumably the gas associated with this dust, are at temperatures of 
several hundred Kelvins, the LTE fractional populations in the low lying 
levels normally observed at radio and millimeter wavelengths would be 
much less than in galactic dense clouds. This cannot be the explanation 
for the non-detection of the CO fundamental band lines, however, as the 
$M$ band spectrum in Fig.~\ref{fig:spectrum} encompasses lines with a 
wide range of lower state energies (corresponding to $J$~=~1--30 in the 
case of the P branch).

The second possibility is that, although there is a large amount of 
molecular material along the line of sight to the nuclear continuum 
source, it is concentrated into clumps that cover only a small fraction 
of the source. Each individual clump produces deep CO lines but as most 
of the continuum source is unobscured, the net effect is much weaker CO 
absorption. Such small cloud volume filling factors have been invoked in 
order to explain the mid-infrared observational properties of Seyfert~1 
and Seyfert~2 galaxies in the unified model \citep[in particular the 
lack of silicate emission in the former and the broad spectral range of 
the far-infrared continuum;][]{nen02}. Small filling factors for toroids 
surrounding AGNs are also predicted by the hydrodynamic simulations of 
\citet{wad07}.

If the detectability of CO absorption is related to the cloud filling 
factor, then for similar total extinction CO should be detected in 
objects with smooth obscuration (i.e., covering the entire nuclear 
continuum source), but not in those with clumpy obscuration. 
\citet{sir08} propose that clumpy and smooth obscuration in AGN can be 
separated on the basis of the strength of the 10 and 18~$\mu$m silicate 
features. Strong CO absorption, in gas that is warm and presumably close 
to the nuclear source, has been detected in several AGN-hosting 
ultraluminous infrared galaxies (ULIRGs) \citep{spo04,geb06,shi07,san08}, 
whose silicate features would classify their dust distribution as 
smooth, whereas in the single Seyfert 2 ULIRG whose spectrum would 
classify it as clumpy, only a weak CO band has been tentatively detected 
\citep{san08}. This may suggest that the strength of the molecular 
absorption in AGN is indeed governed by the clumpiness of the obscuring 
medium, but a stringent test of this hypothesis will require a larger 
sample of objects spanning a wide range of dust geometries.

Nevertheless, the presence in NGC~1068 of the 3.4~$\mu$m feature, whose 
spatial behavior suggests that much of it arises in the same gas as much 
of the silicate feature, close to the AGN, raises immediate 
complications with the clumpy model. That model has the dust and gas 
local to the AGN residing in dense clouds with a small filling factor, 
so that the silicate absorption, which is strong in each clump, is both 
diluted and filled in by emission \citep{nen02}.  The CO absorption 
lines would also be diluted by this arrangement, apparently to 
undetectability in the case of NGC~1068. However, the absence of the 
3.4~$\mu$m band in Galactic dense molecular clouds suggests that in 
NGC~1068 the carrier of that band would not reside in such clumps, and 
thus that a second, more diffuse medium is needed for it (ignoring, for 
the moment the evidence that the carriers of the two dust features are 
mixed). The diffuse material could, for example, be a low-density 
interclump medium. However, the 3.4~$\mu$m band in Galactic diffuse 
clouds is invariably accompanied by a silicate absorption feature. As 
pointed out earlier, the previously estimated value of 
$\tau_{3.4}$/$\tau_{9.7}$ in the diffuse interstellar medium toward the 
Galactic center, when applied to NGC~1068, implies that most or all of 
silicate absorption in NGC~1068 must arise in that material carrying the 
3.4~$\mu$m band.

The by now almost inevitable solution to these problems is the third 
explanation: at 3--13~$\mu$m essentially all of the absorbing material 
in the line of sight has the chemical characteristics of Galactic 
diffuse clouds. In this explanation clumpy absorbing material is not 
needed to account for the lack of detectable CO, because in diffuse 
clouds CO is mostly dissociated, comprising only one percent of the 
carbon. Diffuse clouds do have large molecular components, with H$_2$ 
accounting for typically half of the hydrogen; thus H$_2$ should be 
available to provide the line emission observed by \citet{mul09}. The 
dust grains in diffuse clouds are comprised of a refractory component 
responsible for the 3.4~$\mu$m feature and a silicate component. Gas 
densities in Galactic diffuse clouds are 10--1000~cm$^{-3}$. Densities 
in the inner few tens of parsecs of NGC~1068 might be greater than this. 
However, at distances of only a few tens of parsecs from a luminous AGN 
it may well be possible that somewhat denser gas also has the above 
characteristics.

As discussed earlier, in the Galaxy fairly tight relationships have been 
found between the visual extinction in the diffuse ISM and the depth of 
the 3.4~$\mu$m feature \citep{pen94} for sightlines to the Galactic 
center and to local diffuse clouds. Here we employ the former 
relationship, $A_V$/$\tau_{3.4}$~=~150. Using it we estimate a typical 
visual extinction of 12~mag to the 3.4~$\mu$m continuum-emitting region, 
implying for a normal gas-to-dust ratio that 
$N$(H)~=~2.3~$\times$~10$^{22}$~cm$^{-2}$. Then, assuming that, as in 
Galactic diffuse clouds, one percent of the carbon is in CO, and also 
(conservatively) that the typical continuum surface at 3.4~$\mu$m and 
4.7~$\mu$m are at the same depth, 
$N$(CO)~=~6~$\times$~10$^{16}$~cm$^{-2}$, which is consistent with the 
upper limit of $1 \times$~10$^{17}$~cm$^{-2}$ set by these observations. 
The above column density of hydrogen is much less than the column 
density that attenuates our view of the central engine, 
$N$(H)~$>$~10$^{25}$~cm$^{-2}$ \citep{mat04}, but much of that gas must 
be interior to the dust that is emitting the observed 3--5~$\mu$m 
continuum.

\subsubsection{Summary}

We have found that the 3.4~$\mu$m hydrocarbon and 9.7~$\mu$m silicate 
bands have spatial variations that are similar across the central 
0.2~$\times$~0.8~arcsecond of NGC~1068 and that only a small fraction of 
interstellar carbon in this region can be in the form of CO. The first 
finding indicates that the carriers of the 3.4~$\mu$m and 9.7~$\mu$m 
bands are largely mixed and that significant fractions of each are 
located in material very close to the nucleus. In the Galaxy both 
features are present in diffuse interstellar clouds, but the 3.4~$\mu$m 
band is {\it only} observed there.  The fraction of CO in diffuse clouds 
is small and our observed upper limit in NGC~1068 is consistent with 
such a location. Measurable CO absorption would be present if even a 
modest fraction of the silicate absorption arose in fully molecular 
material. The straightforward conclusion is that {\it the 3-13~$\mu$m 
absorption spectrum of NGC~1068 is produced in material chemically 
similar to the Galactic diffuse interstellar medium.}

We reiterate the significance of the lack of detection of CO and its 
incompatibility with models employing clumps of {\it dense molecular} 
material, even if such material near an AGN is capable of producing the 
3.4~$\mu$m feature.  If the filling factor of such fully molecular 
clumps were low enough to ``dilute" the CO band to non-detection, as 
observed, similar dilutions would occur for the silicate and 3.4~$\mu$m 
absorptions, contradicting the observations.

The adaptive optics measurements of \citet{gra06} show that nuclear 
thermal infrared continuum arises in part from numerous clumps covering 
a few tens of parsecs. If the source of the 3-13~$\mu$m continuum is 
dominated by such clumps, then absorbing material could either be 
located on surfaces of the clumps or be more continuously distributed in 
front of them. However, if a significant portion of the continuum is 
emitted by an extended component, the present observations constrain the 
degree of clumpiness in the absorbing material. In that case a low,
e.g. $<$0.1, filling factor for the absorbing clumps probably would be 
inconsistent with the observed strength of the 3.4~$\mu$m feature, as it 
would imply, e.g., $\tau$(3.4~$\mu$m)~$>$~0.8 in the clumps. 
The maximum observed optical depth of the feature is 0.5, in the deeply 
buried ULIRG IRAS~08572+3915 \citep{geb06}.

\section{Conclusions}
\label{conclude}

The observations reported here bear largely on two important issues: (1) 
the ionization mechanism for the high excitation coronal line emission 
close to the AGN; and (2) the nature of the dusty environment absorbing 
the infrared continuum of the AGN. Regarding the ionized gas, the 
evidence appears to both favor collisional excitation of the bulk of the 
coronal line emission and, in view of the spatial coincidences of the 
various lines, argue against photoionization. Regarding the absorbing 
material, the simplest explanation is that it has much more of the 
characteristics of diffuse clouds than dense clouds (clumped or 
unclumped). The strength of the hydrocarbon feature argues against it 
being distributed in front of the sources of the continuum with a very 
low filling factor relative to them. Because elsewhere it has always 
been found that silicate dust is always present where a 3.4~$\mu$m 
absorption is observed, one can conclude that if the latter is not so 
distributed then neither is the former.

Finally, it should be noted that in NGC~1068 the extended source of the 
3--5~$\mu$m continuum radiation has a characteristic north-south 
dimension of $\sim$20~pc, indicating that the bulk of the continuum 
emission occurs at distances of $\sim$10~pc or more from the central 
engine. This distance is much greater than that estimated under the 
assumption that the temperature of the dust is determined principally by 
the luminosity of the AGN and its distance. Using an AGN luminosity of 
1.5~$\times$~10$^{11}$~L$_{\odot}$ \citep{wei99}, dust grains heated by 
isotropic radiation from the AGN to the observed 3--5~$\mu$m color 
temperatures of $\sim$500~K would lie only $\sim$1~pc from the AGN. 
\citet{gra06} has suggested that transient heating of small grains could 
be responsible for the high temperatures. We speculate that at least in 
the north-south direction mechanical heating of the dust and gas, 
perhaps from the AGN jets, could also be important.

\begin{acknowledgements}

We thank the staff of the United Kingdom Infrared Telescope, which is 
operated by the Joint Astronomy Centre on behalf of the Science and 
Technology Facilities Council of the U.K.  TRG's and REM's research is 
supported by the Gemini Observatory, which is operated by the 
Association of Universities for Research in Astronomy, Inc., under a 
cooperative agreement with the NSF on behalf of the Gemini partnership: 
the National Science Foundation (United States), the Science and 
Technology Facilities Council (United Kingdom), the National Research 
Council (Canada), CONICYT (Chile), the Australian Research Council 
(Australia), Minist\'erio da Ci\'encia e Tecnologia (Brazil) and SECYT 
(Argentina). ARA's research was partially supported by the Brazilian 
funding agency CNPq (311476/2006-6). We are grateful to N. A. Levenson 
for useful discussions and to the anonymous referee for numerous helpful 
comments.

\end{acknowledgements}

\end{document}